\documentclass[journal]{IEEEtran}
\usepackage{amsmath,epsfig,epsf,times,amssymb,booktabs,subfigure}
\usepackage{setspace}
\usepackage{array}
\usepackage{url}
\usepackage{float}
\usepackage{color}
\usepackage{graphicx}
\usepackage{amsthm,cite}
\usepackage{enumerate}
\usepackage{balance}

\theoremstyle{remark}

\begin{document}

\title{Low Probability of Detection Communication: Opportunities and Challenges}
\author{Shihao Yan, Xiangyun Zhou, Jinsong Hu, and Stephen V. Hanly
\thanks{S. Yan and S. V. Hanly are with Macquarie University; X. Zhou is with The Australian National University; J. Hu is with Fuzhou University. The corresponding author of this article is Jinsong Hu.}
}


\maketitle

\vspace{-1cm}

\begin{abstract}
Low probability of detection (LPD) communication has recently emerged as a new transmission technology to address privacy and security in wireless networks. Recent studies have established the fundamental limits of LPD communication in terms of the amount of information bits that can be conveyed from a transmitter to a receiver subject to a constraint on a warden's detection error probability. The established information-theoretic metric enables analytical studies on the design and performance of LPD communication under various channel conditions. In this article, we present the key features of LPD communication and discuss various important design considerations. Firstly, we clarify the differences between LPD communication and the well-known physical-layer security. Then, from an information-theoretic point of view, we discuss the optimal signalling strategies for transmitting the message-carrying signal and artificial-noise signal for LPD communication. Finally, we identify the key challenges in the design of practical LPD communication systems and point out future research directions in this context. This article provides guidelines for designing practical LPD communication strategies in wireless systems and networks.
\end{abstract}


\section{Introduction}

As fifth generation (5G) wireless networks together with the Internet of Things (IoT) are brought to reality, people and organizations become more dependent on wireless devices to share secure and private information (e.g., location, physiological information for e-health).
One impediment to widespread adoption of these technologies are people's concerns about security and privacy of wireless communications.
For example, the exposure of an embedded medical device's transmission may indicate the sickness of a user and may disclose the user's location, which may violate the privacy of the user and is not allowed by the user. To guarantee a strong security or privacy, it is often not sufficient to only protect the content of communications, but it is also required to hide the very existence of wireless transmissions~\cite{bash2015hiding}. Hiding wireless transmissions may also be explicitly desired by government and military bodies (e.g., for a stealth fighter to be able to hide itself from enemies while communicating with its military bases).
However, there is a lack of understanding on the performance limit and enabling technologies for hiding the existence of wireless communications, since current cryptographic and physical-layer security technologies protect only content (i.e., what is transmitted) of wireless communications \cite{yang2015safe,zhao2016physical}.

Hiding wireless transmissions was partially addressed by spread spectrum as the only existing solution in practical use cases. Spread spectrum was invented a century ago with the original purpose of hiding military wireless transmissions (e.g., by spreading transmit power into noise). However, the fundamental performance limit of hiding wireless transmission by spread spectrum has not been fully analyzed, which leads to the fact that there was no clear understanding on when or how often spread spectrum fails to hide wireless transmissions. As such, the amount of covertness achieved by spread spectrum has not been fully revealed.
Due to the unproven and unguaranteed performance, the main use of spread spectrum deviated from hiding wireless transmissions to obtaining high reliability and high data rate in the last two decades. Motivated by the ever-increasing desire of a strong security, cutting-edge research on wireless communication security has called for a rethinking and generalization of spread spectrum (for security purposes) at a more fundamental level, which inspires the emergence of a new security paradigm termed low probability of detection (LPD) communication. Research on LPD communication focuses on the fundamental limits of hiding wireless transmissions, in terms of the amount of information that can be conveyed covertly from a transmitter Alice to a legitimate receiver Bob subject to a specific probability of being detected by a warden Willie~\cite{bash2013limits,bash2015hiding}.

Recent research on LPD communication aims at the potential integration of LPD communication techniques into current and future wireless networks, e.g., 5G wireless networks and IoT, to protect the privacy and security of served users. LPD communication technology is also critical to government and military bodies. For instance, with an increasing cybercrime throughout the world, the police or military units desire to detect communications between any potential cybercriminals, since the presence (not necessarily the content) of such communications offers sufficient alarm for the government bodies or military units to take action. Research in this area will provide a thorough understanding of LPD communication, which can enable the government to (i) enhance national security through foreseeing any future fortunate or devastating impact of this technology on our national cybersecurity, and (ii) understand how to regulate the use of this new technology in future wireless communications.

LPD communication technology has drawn significant research interests since 2013~\cite{bash2013limits}. Existing research in this field can be broadly categorized into three main directions with overlaps. The first category focuses characterizing the performance limits of LPD communication (e.g., \cite{bloch2016covert}), which aims to disclose the number of information bits that can be conveyed with a negligible detection probability. The second one is on encoding schemes to achieve LPD communication, focusing on constructing practical encoding schemes and characterizing the required key size in order to achieve the LPD communication limits (e.g., \cite{wang2016fundamental}). The third category on LPD communication performance enhancements targets at examining possibilities and developing techniques to improve LPD communication performance in realistic environments.
Although the limited preliminary work in each category has provided some initial understanding on LPD communication, many challenging problems and issues have not been addressed. Against this background, in this article we first
discuss the fundamental features of LPD communication and provide a comparison with a closely related technology of physical-layer security.
We will then highlight some existing research results of LPD communication, further identify significant challenging problems and issues in designing practical LPD communication systems, and finally provide our recommendations on future research directions in LPD communication.

The remainder of this article is organized as follows. In Section~\ref{sec:difference}, we introduce the key features of LPD communication and compare it with physical-layer security techniques from the aspects of fundamental frameworks, performance metrics, and mathematical tools.
Some important research results and existing open problems on the optimization of signalling strategies and artificial interference in LPD communication are discussed in Section~\ref{sec:scenario}. Then, we discuss how wireless engineers can apply traditional communication techniques (e.g., modulation, channel coding) to LPD communication in Section~\ref{sec:challenge}. Section~\ref{sec:conclusion} concludes this article.

\section{Exisiting Main Conclusions on LPD Communication and its Key Differences relative to Physical-Layer Security}\label{sec:difference}

In this section, we first present a motivation scenario for LPD communication and some main conclusions in this research direction. Then, we clarify the differences between LPD communication and physical-layer security.

\subsection{A Prisoner's LPD Communication Problem and Spread Spectrum}

\begin{figure}[t]
\centering
\includegraphics[width=0.45\textwidth]{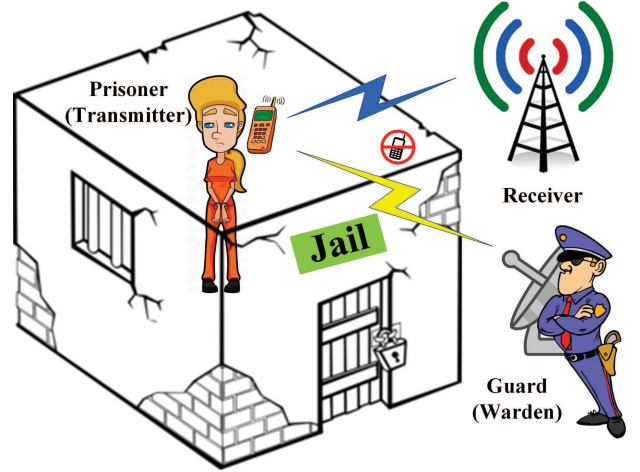}
\caption{A wireless LPD communication scenario, where a prisoner Alice tries to send critical information to a receiver Bob subject to a negligible probability of being detected by the guard Willie.}
\label{fig:fig1}
\end{figure}

In wireless networks, LPD communication problem can be described using a prisoner's communication scenario as shown in Fig.~\ref{fig:fig1}, where a prisoner Alice intends to transmit critical information to a receiver Bob subject to a negligible probability of being detected by the guard Willie. In this scenario, the guard does not care what Alice transmits and he would take off Alice's phone as long as he detects her transmission. As such, from Willie's perspective, he is going to detect whether Alice's wireless transmission occurs or not, which is a binary detection problem. Meanwhile, from Alice's perspective, she has to hide her wireless transmission to Bob in order to avoid losing her phone.

Spread spectrum was actually used widely to hide military wireless communication during World War II. However, the fundamental performance limit of spread spectrum, in terms of the amount of information that can be transferred covertly with a certain covertness level, was unknown. Without such limit analysis, we cannot answer many questions with regard to the fundamental performance limit of spread spectrum in terms of the achieved covertness, e.g., ``\textit{What is the probability of being detected if Alice transmits a certain amount of information (e.g., a photo of a certain size) within a specific period?}'' and ``\textit{Given a maximum tolerance level on the probability of being detected (e.g., $0.5\%$), how much information can Alice transmit with a certain amount of time?}''. Tackling these questions motivated the emergence of LPD communication, aiming at analyzing the fundamental limits of hiding wireless transmissions. LPD communication technology can hide the very existence of wireless transmissions with proven performance and thus can mitigate the threat of discovering the presence of a user or communication to achieve a strong security of ever-increasing demand in wireless civilian and military networks.

\subsection{Square Root Law and Its Extension in LPD Communication}

A square root law was derived in \cite{bash2013limits} by considering additive white Gaussian noise (AWGN) channels for both Bob and Willie, which states that no more than $\mathcal{O}(\sqrt{n})$ bits can be conveyed reliably from Alice to Bob in $n$ channel uses while lower-bounding Willie's detection error probability of this transmission being no less than a specific value $\epsilon$. After \cite{bash2013limits}, the scaling constant of the amount of covert information with respect to $\sqrt{n}$ was characterized for a discrete memoryless channel (DMC) and the AWGN channel in \cite{wang2016fundamental}. The achievability of the square root law normally requires pre-shared secrecies between Alice and Bob prior to Alice's transmission. For DMCs, this key size was shown to be on the order of $\sqrt{n}$ to achieve the square root law regardless of the quality of the channels~\cite{bloch2016covert}. In addition, this pre-shared secret was proven to be unnecessary when the channel quality from Alice to Bob is higher than that from Alice to Willie \cite{bloch2016covert}. We note that the square root law, which leads to zero covert rate, may not hold in all channel models. For example, the work \cite{mukherjee2016covert} showed that non-zero covert rate is achievable in queuing timing channels, where  a sufficiently high rate secret key is available. This also demonstrates the tradeoff between the required key size and the achievable covertness in LPD communication.

\subsection{LPD Communication with the Aid of Artificial Noise}

The performance of LPD communication can be effectively improved by the transmission strategies with the aid of artificial noise (AN)~\cite{sobers2017covert,Khurram2018fullduplex}. It was shown that LPD communication with $\mathcal{O}(n)$ bits over $n$ channel uses could be achieved when a uniformly distributed ``jammer'' is present to help the transmitter Alice in AWGN and block fading channels \cite{sobers2017covert}. Besides external jammers, AN transmitted by a full-duplex receiver Bob can enable LPD communication and significantly improve its performance~\cite{Khurram2018fullduplex}. Specifically, it was shown that the transmission of AN with varying power, although causing self-interference at Bob, provides the opportunity of achieving covertness without other uncertainties at Willie in fading channels, where it was demonstrated that the transmit power range of AN should be managed carefully in order to enhance LPD communication.

\subsection{LPD Communication in Random Wireless Networks}

The ultimate goal of LPD communication is to achieve shadow wireless networks~\cite{bash2015hiding}. Preliminary results towards this goal have been achieved in the context of random wireless networks. LPD communication with a Poisson field of interferers was analyzed in \cite{Biao2018Poisson}, which drew interesting conclusions with regard to the impact of interferers. Firstly, it showed that the density and transmit power of the concurrent interferers do not affect the covert throughput when the network stays in the interference-limited regime. When the interference is proportional to the receiver noise, the covert throughput increases with the density and transmit power of the interferers. In addition to interferers, the work of \cite{zheng2019multi} considered randomly located wardens in LPD communication with centralized and distributed transmit antennas at Alice. It drew a similar conclusion as \cite{Biao2018Poisson}, i.e., the maximum covert throughput is invariant to the density or transmit power of interferers regardless of the transmit antenna number or locations (centralized or distributed).


\subsection{Multi-Hop LPD Communication}

Another preliminary step towards achieving shadow wireless networks is the research work on multi-hop routing in LPD communication \cite{shei2018multi}. The covertness requirement in LPD communication highly limits its communication range and thus multi-hop communication is desired in shadow wireless networks. Against this background, \cite{shei2018multi} considered multi-hop routing with multiple relays to achieve a long-distance LPD communication from Alice to Bob. The maximum throughput and minimum end-to-end delay in the presence of multiple collaborating wardens were achieved in two cases, where a single key or multiple independent keys were used at the relays, respectively. The results in \cite{shei2018multi} show that multi-hop transmission significantly improve the performance of LPD communication relative to the single-hop transmission. Furthermore, their results showed that the case with multiple independent keys outperforms that with a single key, which indicates the tradeoff between system complexity (and overhead cost) and the LPD communication performance.

\subsection{Delay-Intolerant LPD Communication}

Most works in the literature of LPD communication consider the number of channel uses being large or asymptotically infinite, which leads to large communication delay. Motivated by some delay-intolerant applications, the work \cite{Shihao2018Delay} investigated delay-intolerant LPD communication, which showed that under a specific delay constraint the transmission should occur within all the available channel uses in order to maximize the effective amount of information that can be conveyed covertly from Alice to Bob. Random transmit power was also considered in \cite{Shihao2018Delay}, which drew another general conclusion in LPD communication. That is, as long as the uncertainty is not zero, it can be increases by some strategy, which turns out to improve the performance of LPD communication. This conclusion was also confirmed by the work \cite{shu2019delay}, which showed that AN transmitted with a fixed power (not a varying one) can enhance the delay-intolerant LPD communication.

\subsection{Differences between LPD Communication and Physical-Layer Security}

\begin{figure}[t]
\centering
\includegraphics[width=0.48\textwidth]{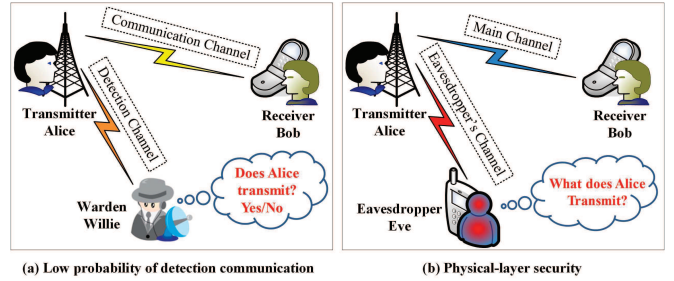}
\caption{System models of (a) low probability of detection (LPD) communication and (b) physical-layer security.}
\label{fig:fig2}
\end{figure}

Researchers from the physical-layer security community have started to see LPD communication as a closely related technology for achieving security. Nevertheless, there are fundamental differences between LPD communication and physical-layer security. To clarify these differences, we first present their fundamental features, as illustrated in Fig.~\ref{fig:fig2}. From this figure, we can see that the main difference between LPD communication and physical-layer security is in the nature of the malicious user. This malicious user in LPD communication is the warden Willie, who cares whether the transmitter Alice transmits information to the legitimate receiver Bob. This is also the reason that it is a communication channel from Alice to Bob, while it is a detection channel from Alice to Willie.
Meanwhile, the malicious user in physical-layer security is the eavesdropper Eve, who intends to know what Alice transmits to Bob~\cite{yang2015safe,zhao2016physical}. In physical-layer security, the channel from Alice to Bob is the main channel and the channel from Alice to Eve is the eavesdropper's channel, which both are for communications. Mathematically, Willie is dealing with a binary detection problem in LPD communication, while Eve is facing a communication problem in physical-layer security.

\section{On the Information-Theoretic Optimality of Signalling \\in LPD Communication}\label{sec:scenario}

In this section, from an information-theoretic point of view, we first discuss the optimality of Gaussian signalling in LPD communication. Then, we identify challenging research problems with regard to the optimality of the AN or interference when such noise or interference is controllable.

\subsection{Optimality of Gaussian Signalling in LPD Communication}
In communication theory, it is known that Gaussian signalling (i.e., the transmitted signal follows a normal distribution) is optimal for traditional point-to-point communication in terms of maximizing the mutual information between the input and output in AWGN channels. Following this conclusion, a straightforward question is whether Gaussian signalling is optimal in LPD communication.
Tackling this question with Willie's minimum detection error probability being no less than a specific value as the covertness constraint is very challenging, since this error probability is mathematically intractable before determining the distribution of Alice's transmitted signal \cite{yan2018gaussian}.
We note that Willie's minimum detection error probability equals to one minus the total variation $\mathcal{V}_T(p_{_0}, p_{_1})$ between the likelihood function $p_{_0}$ under the null hypothesis and the likelihood function $p_{_1}$ under the alternative hypothesis. However, $\mathcal{V}_T(p_{_0}, p_{_1})$ is also mathematically intractable in most scenarios~\cite{bash2013limits}. Meanwhile, Kullback-Leibler (KL) divergence offers tight bounds on $\mathcal{V}_T(p_{_0}, p_{_1})$ and this KL divergence can be expressed in analytical expressions. Thus, the KL divergence has been widely used in the literature of LPD communication to set tractable covertness constraints~\cite{yan2018gaussian}. Due to the asymmetry property of KL divergence, we can use either $\mathcal{D}({p_{_0}||p_{_1}})$ or $\mathcal{D}({p_{_1}||p_{_0}})$ to determine a covertness constraint, where $\mathcal{D}({p_{_0}||p_{_1}})$ is the KL divergence $p_{_0}$ to $p_{_1}$ and $\mathcal{D}({p_{_1}||p_{_0}})$ is the KL divergence from $p_{_1}$ to $p_{_0}$.
It is shown that Gaussian signalling is optimal in terms of maximizing the mutual information between the input and output of the communication channel subject to the covertness constraint determined by $\mathcal{D}({p_{_1}||p_{_0}})$. This is due to the fact that the normal $p_{_1}$ can simultaneously maximize the considered mutual information and minimizing $\mathcal{D}({p_{_1}||p_{_0}})$ as proved in \cite{yan2018gaussian}. However, we note that Gaussian signalling is not optimal when the covertness constraint is determined by $\mathcal{D}({p_{_0}||p_{_1}})$, since the normal $p_{_1}$ cannot minimize $\mathcal{D}({p_{_0}||p_{_1}})$ while maximizing the mutual information. As shown in \cite{yan2018gaussian}, a skew-normal $p_{_1}$ can achieve a lower $\mathcal{D}({p_{_0}||p_{_1}})$, which leads to the result that a skew-normal $p_{_1}$ can possibly outperform the normal $p_{_1}$ in the context of LPD communication. Following this, a challenging and information-theoretic research problem in LPD communication is ``\textit{What is the optimal signalling strategy in the LPD communication with the covertness constraint determined by $\mathcal{D}({p_{_0}||p_{_1}})$?}''.

\begin{figure}[t]
\centering
\includegraphics[width=0.48\textwidth]{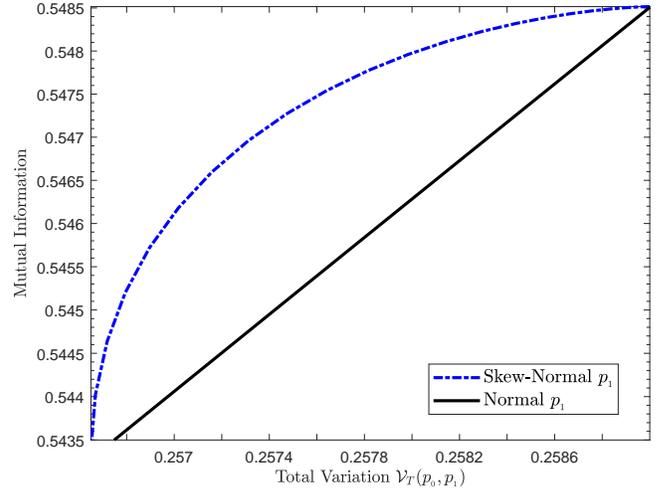}
\caption{Mutual information between the input and output of the communication channel versus $\mathcal{V}_T(p_{_0}, p_{_1})$ in LPD communication, where AWGN at Bob and Willie are independent and identically distributed.}
\label{fig:fig3}
\end{figure}

In \cite{yan2018gaussian}, it is also numerically demonstrated that Gaussian signalling is still not optimal when the covertness constraint is actually based on $\mathcal{V}_T(p_{_0}, p_{_1})$, which is also shown in Fig.~\ref{fig:fig3}. In this figure, we observe that for a given value of $\mathcal{V}_T(p_{_0}, p_{_1})$ the skew-normal $p_{_1}$ can achieve a higher mutual information. This is due to the fact that the skew-normal $p_{_1}$ can lead to a smaller $\mathcal{V}_T(p_{_0}, p_{_1})$ than the normal $p_{_1}$, although skew-normal $p_{_1}$ cannot achieve a higher mutual information than the normal $p_{_1}$. This observation indicates that Gaussian signalling is not optimal in practical LPD communication. Against this background, examining the optimal signal strategy in practical LPD communication deserves much near future research effort.


Based on the above discussions, we know that Gaussian signalling is not optimal in practical LPD communication, since it cannot optimally hide the transmitted signal from the Warden Willie, although it achieves the best communication performance from the transmitter Alice to the receiver Bob. This provides two possibilities on the optimal signalling strategy in LPD communication. The first one is that the optimal signalling may be the one that can optimally hide the transmitted signal in terms of maximizing the detection error probability at Willie. The other one is that the optimal signalling can be the tradeoff between the one optimally hides transmitted signal and the one optimizes the communication from Alice to Bob. These two possibilities deserve future ongoing research effort, which will facilitate clarifying how the transmit signal is optimally hidden while carrying the maximum useful information.

\subsection{Optimality of Artificial Noise and Interference in LPD Communication}



Considering finite channel uses, AN or interference with a fixed power can enhance LPD communication (e.g., \cite{Shihao2018Delay,shu2019delay}). In this case, the distributions of the AN and interference can be optimized to maximize the communication channel mutual information subject to the performance limits of the detection channel. This optimization is a challenging research problem, since general expressions for the distributions of Bob and Willie's received signals are hard or infeasible to achieve for a given distribution of the AN or interference. Following the work \cite{yan2018gaussian}, we note that calculus of variations may serve as the main mathematical tools for solving such optimization problems. In addition to directly optimizing AN or interference, the impact of information-theoretic (IT) coding with finite blocklength should be considered in this optimization, since IT coding not only affects the detection performance at the warden Willie, but also has direct impact on the decoding error probability at the receiver Bob.

In the case with infinite channel uses, the transmit power of such AN or interference should be randomized in order to enhance LPD communication~\cite{Khurram2018fullduplex}. Along this direction, future research effort should be on identifying the optimal distribution of the transmit power for the AN or interference subject to practical constraints (e.g., maximum transmit power constraint) in this case. In existing works, the transmit power distribution of AN or interference is typically set to follow a uniform distribution as an arbitrary choice (e.g., \cite{Khurram2018fullduplex}).


\section{Practical Design Challenges and Open Research Problems in LPD Communication}\label{sec:challenge}

LPD communication is to hide the transmission from the warden Willie while conveying as much information as possible from the transmitter Alice to the receiver Bob. Relative to traditional wireless communications, the new aspect of LPD communication is to hide wireless transmission from Willie, i.e., the covertness constraint. The principle of achieving this covertness is to make the detection of Alice's transmission hard for Willie. To this end, in designing practical LPD communication we have to keep non-zero uncertainty in Willie's binary detection and increase this uncertainty by all means. In the following, we clarify some challenges in designing practical LPD communication systems and present potential future research directions in LPD communication.

\subsection{Modulation}

\begin{figure}[t]
\centering
\includegraphics[width=0.48\textwidth]{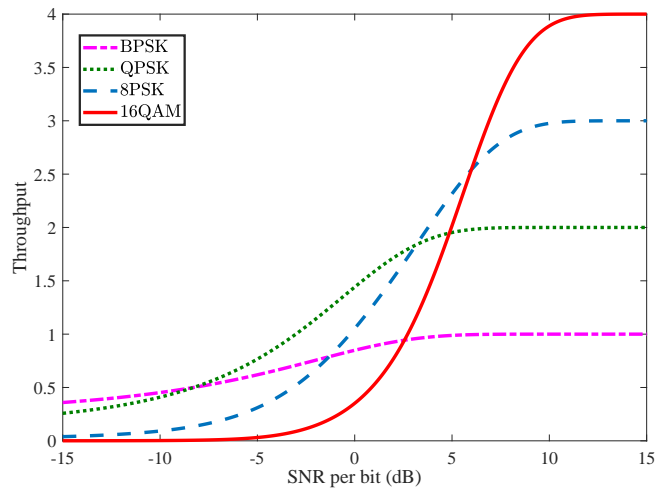}
\caption{Throughput achieved by four different modulation schemes, i.e., BPSK, QPSK, 8PSK, and 16QAM, versus bit SNR.}
\label{fig:fig4}
\end{figure}

Achieving LPD communication usually means that Alice's transmit power needs to be small, thus Bob's signal reception experiences low signal-to-noise ratio (SNR). As such, the detection and throughput performance of different modulation schemes in the low SNR regime is of great importance.  In the low SNR regime, low-order modulation performs better than high-order modulation. Intuitively, this is due to the fact that in the low SNR regime a low-order modulation (corresponding to the low information rate without considering channel coding) can provide a higher reliability than a high-order modulation.
To confirm this, in Fig.~\ref{fig:fig4} we plot the achieved throughput of four different modulation schemes, i.e., Binary Phase Shift Keying (BPSK), Quadrature PSK (QPSK), 8PSK, and  16-Quadrature Amplitude Modulation (16QAM), versus SNR per bit, where the throughput is defined as the product of the rate of each modulation scheme and the corresponding bit success rate (that is, one minus bit error rate). In this figure, we observe that modulation schemes with lower modulation orders achieve favourable performance in the low SNR regime. On the other hand, for detection at Willie, as the modulation order becomes higher, the received signal statistically follows a mixture distribution with more components. As per the central limit theorem, this mixture distribution approaches to a Gaussian distribution as the number of components increases, which generally leads to a higher detection error probability at Willie.
As such,  a modulation with a higher order for Alice's transmission to Bob makes it harder for Willie to detect this transmission. Following this and considering the performance of different modulation schemes in the low SNR regime, there exists an optimal modulation scheme for LDP communications under some specific conditions. In this context, determining these conditions and the corresponding optimal modulation scheme are future research directions of significant importance. We note that these conditions and the choice of the optimal modulation scheme will be highly dependent on the covertness requirements.


\subsection{Channel Coding}

Apart from the nontrivial tradeoff in modulation order, the use of channel coding in Alice's transmission can also be a double-edged sword. It is well known that channel coding can improve communication reliability by adding extra redundancy into transmitted information bits. The added redundancy also increases the number of received symbols or samples at Willie (i.e., increasing the chance for Willie to detect Alice's transmission) when Alice transmits a certain amount of information. As such, examining the conditions under which channel coding enhances the performance of LPD communication deserves future research effort. Furthermore, the relative performance of different channel coding schemes in LPD communication may differ from that in traditional wireless communications, which also serves as another future research direction in this context. Finally, we note that channel coding with finite blocklength is different from that with sufficiently large blocklength, which serves as an independent research direction in the context of LPD communication. This is due to the fact that with finite blocklength the decoding error probability at the receiver Bob is not negligible anymore and the coding strategy should simultaneously consider Bob's decoding error probability and Willie's detection performance. This research direction is challenging but has the most practical impact, since the communication delay (and thus the blocklength) is finite in any practical scenario.

\subsection{Channel State Information}


In LPD communication, a question to answer before conducting channel estimation is whether channel state information (CSI) should be estimated at all. In other words, comparison between a coherent communication (which requires CSI at the receiver) and a fully non-coherent communication (which does not require CSI at the receiver) in the context of LPD communication is needed.
As such, one needs to analyze the performance of non-coherent and coherent communications in the context of LPD communication. Non-coherent transmission (e.g., differential phase shift keying) does not require channel training and thus it does not give any reference signal for Willie to perform detection, which leads to the possibility that the non-coherent transmission outperforms the coherent one in the context of LPD communication. In order to reveal the performance of coherent communication in LPD communication, one has to develop novel and robust channel training schemes to enable coherent covert transmissions, where optimal resource allocation should be considered.

\subsection{Multi-Hop Communications}

\begin{figure}[t]
\centering
\includegraphics[width=0.48\textwidth]{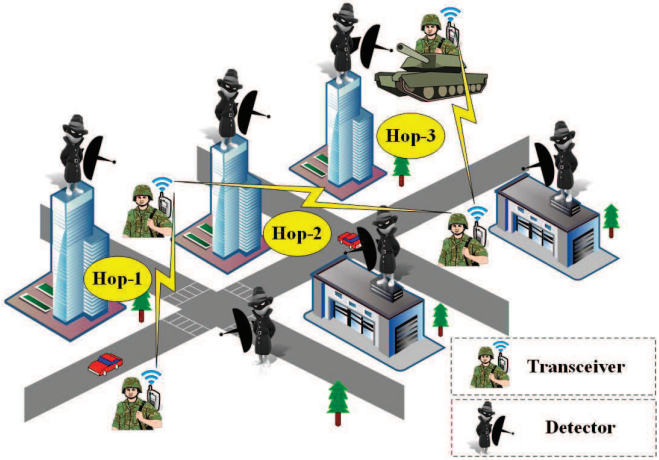}
\caption{An application scenario of multi-hop low probability of detection (LPD) communication.}
\label{fig:fig5}
\end{figure}

The low-power nature of LPD communication significantly restricts the communication range. Therefore, multi-hop communication is essential in many applications of LPD communication where the end-to-end communication distance is large \cite{bash2015hiding,shei2018multi}. In Fig.~\ref{fig:fig5}, we show an application scenario of multi-hop LPD communication, where single-hop communications cannot achieve enough covertness. This is due to the fact that the communication distance is large and thus the transmit power should be high to support such a long-distance communication. A high transmit power makes it easy for a nearby detector to detect this communication and thus the single-hop communication cannot achieve much covertness.


The tradeoff between the communication distance of each hop and the number of hops in achieving long-distance LPD communication has never been analyzed. Specifically, whether more short-distance hops or fewer long-distance hops are desired in LPD communication is not known. In addition to the tradeoff between the number of hops and each hop distance, a novel framework to study the scalability of multi-hop LPD communication is needed to establish covert and inherently secure large-scale wireless networks. In this context, another future research direction is to design practical transmission strategies to achieve acceptable scaling behavior of LPD communication performance with respect to the density of legitimate transceivers or wardens, taking into account the realistic spatial characteristics of wireless channels.

\section{Conclusions}\label{sec:conclusion}

The LPD communication technology can be used to safeguard commercial, government, and military wireless networks by hiding the very existence of wireless transmissions, which can significantly boost society's confidence in exchanging confidential data through wireless communications anytime, anywhere.
In this article, we first present some main conclusions in LPD communication and clarified its difference relative to physical-layer security. We also discussed the optimality of transmitted information signals, AN, and interference in LPD communication.
Our discussion showed that Gaussian signalling, which is the optimal signalling distribution for conventional communications in AWGN channels, is no longer optimal in LPD communication. Finally, the design challenges and open problems identified in this article, including aspects of modulation, channel coding, channel estimation, and multi-hop communications, provided useful references for future research directions in LPD communication.


\vspace{-1cm}

\balance

\begin{IEEEbiographynophoto}\\
Shihao Yan [M'15] (shihao.yan@mq.edu.au) received the Ph.D. degree in Electrical Engineering from The University of New South Wales, Sydney, Australia, in 2015. From 2015 to 2017, he was a Postdoctoral Research Fellow in the Research School of Engineering, The Australian National University, Canberra, Australia. He is currently a University Research Fellow in the School of Engineering, Macquarie University, Sydney, Australia.
His current research interests are in the areas of wireless communications and statistical signal processing, including physical layer security, covert communications, and location spoofing detection.
\end{IEEEbiographynophoto}


\begin{IEEEbiographynophoto}{}
Xiangyun Zhou [SM'17] (xiangyun.zhou@anu.edu.au) is an Associate Professor at the Australian National University. His research interests are in the fields of communication theory and wireless networks. He has served as an Editor for IEEE TRANSACTIONS ON WIRELESS COMMUNICATIONS, IEEE WIRELESS COMMUNICATIONS LETTERS and IEEE COMMUNICATIONS LETTERS. He is a recipient of the Best Paper Award at ICC'11 and IEEE ComSoc Asia-Pacific Outstanding Paper Award in 2016. He was named the Best Young Researcher in the Asia-Pacific Region in 2017 by IEEE ComSoc Asia-Pacific Board.
\end{IEEEbiographynophoto}

\vspace{-1cm}

\begin{IEEEbiographynophoto}\\
Jinsong Hu [M'19] (jinsong.hu@fzu.edu.cn) received the B.S. degree and Ph.D. degree from the School of Electronic and Optical Engineering, Nanjing University of Science and Technology, Nanjing, China in 2013 and 2018, respectively. From 2017 to 2018, he was a Visiting Ph.D. Student with the Research School of Engineering, Australian National University, Canberra, ACT, Australia. He is currently a Lecturer with the College of Physics and Information Engineering, Fuzhou University, Fuzhou, China. His research interests include array signal processing, covert communications, and physical layer security.
\end{IEEEbiographynophoto}

\vspace{-1cm}

\begin{IEEEbiographynophoto}\\
Stephen V. Hanly [M'96-SM'15-F'17] (stephen.hanly@mq.edu.au) received his B.Sc. and M.Sc. from the University of Western Australia, and the Ph.D. degree in mathematics from Cambridge University. He has been a Post-doctoral member of technical staff at AT\&T Bell Laboratories in Murray Hill, New Jersey, and on the research and teaching staff at the University of Melbourne and the National University of Singapore. He is now a Professor at Macquarie University. He has undertaken IEEE Journal editorial roles and has been a General or Technical co-chair of a number of IEEE conferences. He is a member of the Board of Governors of the IEEE Information Theory Society, and he is a Fellow of the IEEE.
\end{IEEEbiographynophoto}

\end{document}